# Thermodynamically stable room-temperature superconductors in Li-Na hydrides under high pressures

Decheng An[1], Defang Duan[1,*], Zihan Zhang[1], Qiwen Jiang[1], Hao Song[2], Tian Cui[2,1,*]

[1]*State Key Laboratory of Superhard Materials, College of Physics, Jilin University, Changchun 130012, China*

[2]*Institute of High Pressure Physics, School of Physical Science and Technology, Ningbo University, Ningbo, 315211, China*

*Corresponding Authors, duandf@jlu.edu.cn, cuitian@nbu.edu.cn

**Abstract**

Room-temperature superconductivity has been a long-standing goal for scientific progress and human development. Thermodynamic stability is a prerequisite for material synthesis and application. Here, we perform a combination of high-throughput screening and structural search and uncover two thermodynamically stable room-temperature superconductors, $Fd$-$3m$-$Li_2NaH_{17}$ and $Pm$-$3n$-$LiNa_3H_{23}$, exhibiting extraordinary critical temperature of 340 K at 300 GPa and 310 K at 350 GPa, respectively. $Li_2NaH_{17}$ possesses the highest $T_c$ among all the thermodynamically stable ternary hydrides hitherto found. The dominated H density of states at the Fermi level and the strong Fermi surface nesting are favorable for the emergence of room-temperature superconductivity. Their excellent superconducting properties help us understand the mechanism of room-temperature superconductivity and find new room-temperature superconductors. Interestingly, the structures of $LiNa_3H_{23}$ and $Li_2NaH_{17}$ equal to the identified type-I and II clathrate geometry. Our results provide a structural reference and theoretical guidance for later experimental structure determination and theoretical search for high temperature superconductors.

**Introduction**

Owing to great potential value of room-temperature superconductivity in civil and industrial application, the pursuit of room-temperature superconductivity has become one of hot issues in condensed matter physics and materials science. Hydrogen, as the first element in periodic table of elements, possesses the lightest mass and simplest electronic structure. Due to high Debye temperature and strong electron-phonon coupling, metallic hydrogen has been considered as a remarkable superconductor with a $T_c$ of 100-760 K[1,2]. However, the insulator solid hydrogen would be metallized above 500 GPa[3–5]. It's a huge challenge to achieve such an extreme pressure in experimental techniques. The dilemma has been solved until Ashcroft proposed "chemical precompression"[6] in hydrogen-rich materials, which could effectively reduce the metallization pressure of solid hydrogen by doping other elements in hydrogen. Within the framework of this theory, a sequence of pressure-induced hydrogen-based superconductors have been predicted and synthesized successfully.[7–9] The theory-oriented findings of $H_3S$ with a record $T_c$ of 203 K at 155 GPa, where S atoms and H atoms form covalent networks, promoted the development of hydrogen-based superconductors[10–12]. A series of clathrate hydrides, $LaH_{10}$, $YH_9$, $YH_6$, and $CaH_6$, exhibiting high $T_c$ above 200 K at high pressures, have been theoretically predicted and experimentally synthesized[13–19].

The future of hydride superconductivity exploration is achieving a right equilibrium between the thermodynamical stability and superconducting critical temperature. Theoretical prediction has proved $Fd$-$3m$-$Li_2MgH_{16}$[20] with a $T_c$ of 351 K at 300 GPa and 473 K at 250 GPa. It updates the upper limit on the hydride $T_c$ by theoretical calculation and inspires researchers to explore room-temperature superconductors. However, $Fd$-$3m$-$Li_2MgH_{16}$ is a thermodynamically metastable phase and has yet to be synthesized. $LaBeH_8$[21] in a $Fm$-$3m$-$AXH_8$ structure has been shown to possess a $T_c$ of 183 K at a much lower pressure of 20 GPa. It provides a strategy for finding high $T_c$ at lower pressures. Recently, thermodynamically stable clathrate hydrides $MH_{18}$[22] (M = rare-earth or actinide atom) with $H_{36}$ clathrate cage have been discovered, in which $Fddd$-$CeH_{18}$ has a $T_c$ of 330 K at 350 GPa. Looking for thermodynamically stable room-temperature superconductors at high pressures and high temperature superconductors at low pressures is two goals of exploration.

For some hydrides that have been theoretically predicted or experimentally synthesized, we could find the same or similar crystal structures in non-hydrides at ambient pressure. The structure of $Im$-$3m$-$H_3S$[10] is similar with $Im$-$3m$-$SF_6$[23]. The $Im$-$3m$-Ca/YH$_6$[13,14,24,25] and $Fm$-$3m$-La/YH$_{10}$[13,14] adopt the known solidate and zeolite structures. The structures of experimental synthesis $Pm$-$3n$-Ba/La/Lu/Eu$_4$H$_{23}$[26–29] and theoretical prediction $Fd$-$3m$-Li$_2$La/YH$_{17}$[30] are equal to the so-called type-I and type-II silicon clathrate structures[31], respectively. This type-I clathrate structure is the Weaire−Phelan foam structure, the optimal solution to the Kelvin problem so far[32]. In type-I and type-II silicon clathrates, Si-Si forms $sp^3$ hybridized bonds[33]. Many clathrate hydrides are similar with $sp^3$-boned clathrate structures, e.g., H$_{24}$ cage in CaH$_6$[18] and H$_{32}$ cage in LaH$_{10}$[13,14] can also be found in Sr(BC)$_3$[34] and La(BN)$_5$[35], respectively. Though H lacks $p$ orbitals, H atoms form clathrate cages like $sp^3$-bonded clathrates. So far, although type-I and type-II clathrate structures have been discovered in hydrides, they have not attracted much attention or been studied extensively. Such highly symmetric structures provide excellent structural templates for studying hydride superconductivity.

To obtain hydrides with excellent superconducting properties, in addition to excellent structures, it is also necessary to choose right atoms or right atomic radii. Especially for the clathrate hydrides, the appropriate atomic radius selection is very important. Because Y and Ce have similar atomic radii with La[36], solid solution hydrogen-based superconductors in La-Y-H[37] and La-Ce-H[38–40] system have been synthesized. $Fm$-$3m$-(La,Y)H$_{10}$[37] has a $T_c$ of 253 K at 183 GPa. $P6_3/mmc$-(La,Ce)H$_9$[38] and $Fm$-$3m$-La$_{0.5}$Ce$_{0.5}$H$_{10}$[40] exhibit $T_c$ of 148-178 K at 97-172 GPa and 175 K at 155 GPa, respectively. Their $T_c$s exceed those of binary hydrides in the same pressure range.

Although the excellent crystal structures and appropriate atomic radius selections are essential for hydride superconductivity, it is extremely difficult to obtain a thermodynamically stable hydride with a high $T_c$ under these conditions. In most cases, most of previous searches for hydrogen-based superconductors rely on the chemical and physical intuition and perform an extensive structure search in a certain system. It is not only expensive but also possible to get the unwanted results. Precise and efficient calculation is very essential. Therefore, the combination of structural screening based on the high-throughput computation and structure search combined with the first-principles software is a good choice. Type-I and type-II clathrate structures have been discovered but not been widely and deeply studied in hydrides. There is only one Wyckoff position difference of $Fd$-$3m$-$Li_2MgH_{16}$ with type-II clathrate structures. Whether a thermodynamically stable room-temperature superconductor like that of $Li_2MgH_{16}$ can also exist in this structure. Some alkali metals and alkaline-earth metals as guest atoms are captured by type-I silicon-host clathrate structures ($A_8Si_{46}$, A = guest atom) with a low-temperature superconductivity have been reported.[41,42] So, we choose $Li_2MgH_{16}$-type, type-I and type-II clathrate structures as the original structures. These cages are stuffed with different radius metal atoms depending on their size. We obtained 12 nearly dynamically stable structures at 300 GPa by high-throughput calculation. Surprisingly, we found that $Li_2NaH_{16}$, $Li_2NaH_{17}$ and $LiNa_3H_{23}$ belong to same system are nearly dynamically stable at 300 GPa. And their H DOS are dominant at the Fermi level. It is very beneficial to obtain high $T_c$ superconductivity.

In this paper, we focused on Li-Na-H ternary system and preformed a wide structure search. We discoveried three thermodynamically stable clathrate hydrides, $Li_2NaH_{16}$, $Li_2NaH_{17}$ and $LiNa_3H_{23}$. Among them, $Li_2NaH_{17}$ and $LiNa_3H_{23}$ are extraordinary room-temperature superconductors with $T_c$ of 357 K at 220 GPa and 323 K at 320 GPa, respectively. $Li_2NaH_{16}$ also has a high $T_c$ of 258 K at 230 GPa. Remarkably, $Li_2NaH_{17}$ possesses the highest $T_c$ of 340 K at 300 GPa among all the thermodynamically stable hydrides. $Li_2NaH_{17}$ can be considered as an extra hydrogen atom introduced into $Li_2NaH_{16}$ in composition. The introduction of the extra hydrogen atom leads to an obvious increase in the number of H electronic states at the Fermi level and the intensity of Fermi surface nesting, which enhances the $T_c$ by nearly 120

K. LiNa$_3$H$_{23}$ and Li$_2$NaH$_{17}$ adopt the type-I and II clathrate structures, these kinds of clathrate structures make it possible to investigate room-temperature superconductors. Our results reveal the effect of electronic structures on superconductivity of hydrides and provide a reference to find high-temperature or even room-temperature superconductors in type-I and II clathrate structures.

**Result and discussion**

Firstly, we performed a high-throughput calculation on Li$_2$MgH$_{16}$-type, type-I and type-II clathrate structures by PHONONPY code[43] to confirm the dynamical stability of structures, small radius atom A (A=Li, Be) and big radius atom B (B=Na, K, Rb, Cs, Mg, Ca, Sr, Ba) are selected to occupy small cages and big cages[36], respectively. The Li$_2$MgH$_{16}$-type and type-II clathrate structures adopt the same space group of *Fd*-3*m* and have only one Wyckoff position difference. The 28-vertex cages form a diamond network, and the rest space is filled with 18-vertex or 20-vertex cages in Li$_2$MgH$_{16}$-type and type-II clathrate structures. The type-I clathrate structure with *Pm*-3*n* symmetry consists of 24-vertex tetrakaidecahedra cages and 20-vertex pentagonal dodecahedra cages. The 24-vertex cages arrange along three perpendicular directions in space and the rest space is filled with 20-vertex cage. We obtained 12 nearly dynamically stable structures through high-throughput calculation. Next, we calculated the projected electronic density of states for these structures. Significantly, Li$_2$NaH$_{16}$, Li$_2$NaH$_{17}$ and LiNa$_3$H$_{23}$ are with abundant H electronic states near the Fermi level, inspiring us to further exploration. Therefore, we performed an extensive structure search and constructed the ternary convex hull of Li-Na-H system at 350 GPa, and found three thermodynamically stable clathrate hydrides, Li$_2$NaH$_{16}$, Li$_2$NaH$_{17}$ and LiNa$_3$H$_{23}$, as shown in Figure 1. Furthermore, their thermodynamically stable pressure range were determined by combining enthalpy difference calculations. The Li$_2$NaH$_{16}$, Li$_2$NaH$_{17}$ and LiNa$_3$H$_{23}$ become thermodynamically stable about 250 GPa, 255 GPa and 335 GPa, respectively (see Figure S2-4).

To examine potential superconductivity in these structures, we calculated the logarithmic average phonon frequency and electron-phonon coupling (EPC) parameters based on linear response theory, as shown in Table SI. It is heartening to note that the resulting EPC parameters $\lambda$ of $Li_2NaH_{16}$, $Li_2NaH_{17}$ and $LiNa_3H_{23}$ are quite large and reach 1.56, 2.44 and 2.69 at 300 GPa, 300 GPa and 350 GPa, respectively. The bending modes of atomic hydrogen (800 – 2500 cm$^{-1}$) makes a major contribution to $\lambda$, they contribute nearly 81%, 79% and 65% of total $\lambda$ in $Li_2NaH_{16}$, $Li_2NaH_{17}$, and $LiNa_3H_{23}$, respectively. Generally, the Eliashberg equations could provide a reliable $T_c$ for strong EPC ($\lambda > 1.5$)[44]. Therefore, we calculated their $T_c$s using Eliashberg equations with typical Coulomb pseudopotential parameters $\mu^*$ from 0.1 to 0.13, listed in Table SI. Encouragingly, these thermodynamically stable hydrides exhibit extraordinary room-temperature superconductivity. The $T_c$ of $Li_2NaH_{17}$ and $LiNa_3H_{23}$ is estimated to be 340 K (321 K with $\mu^* = 0.13$) at 300 GPa and 310 K (291 K with $\mu^* = 0.13$) at 350 GPa, respectively. Notably, $Li_2NaH_{17}$ is thermodynamically stable at 300 GPa and it possesses the highest $T_c$ in all known thermodynamically stable hydrides. As the pressure decreases, the calculated $T_c$ of $Li_2NaH_{17}$ and $LiNa_3H_{23}$ increases to 357 K (340 K with $\mu^* = 0.13$) at 220 GPa and 323 K (302 K with $\mu^* = 0.13$) at 320 GPa, respectively. For $Li_2NaH_{16}$, the calculated $T_c$ is 221 K (201 K with $\mu^* = 0.13$) at 300 GPa. When the pressure drops to 230 GPa, the $\lambda$ increases to 2.69, and the calculated $T_c$ increases to 258 K. The $T_c$ of $Li_2NaH_{16}$, $Li_2NaH_{17}$ and $LiNa_3H_{23}$ increases gradually with decreasing pressures, as shown in Figure 2.

Fermi surface nesting plays a role in superconductivity[45]. We calculated total Fermi surface nesting of $Li_2NaH_{17}$. There is a strong Fermi surface nesting near $X$ point along $\Gamma$-$X$ path. To determine the contribution of each band to Fermi surface nesting, we calculated Fermi surface nesting from different bands to specified band (see Figure 3). We could find band n = 3 makes a main contribution to Fermi surface nesting than other bands. The intensity of Fermi surface nesting depends on the shapes of Fermi surfaces. Several Fermi surfaces paralleling to Fermi surface n = 3 result the strong Fermi surface nesting, which contributes to room-temperature superconductivity.

The calculated band structures and the electronic density of states of $Li_2NaH_{16}$, $Li_2NaH_{17}$ and $LiNa_3H_{23}$ are shown in Figure 4. For $Li_2NaH_{16}$, there are three bands (denoted as n =1, 2, and 3) crossing the Fermi level. The band n =1 forms a flat band at $\Gamma$ point. The band n = 2 exists an electron-like band at $\Gamma$ point. In the case of $Li_2NaH_{17}$, four bands (denoted n =1, 2, 3 and 4) cross the Fermi level. The band structures exhibit electronlike and hole-like bands around Fermi level. The band n=2 along $L$-$\Gamma$ path just above Fermi level exists a hole-like band, resulting a van Hove singularity near the Fermi level. For $LiNa_3H_{23}$, there are five bands (denoted n = 1, 2, 3, 4 and 5) crossing Fermi level at 350 GPa. The band n = 4 exists electron-like and hole-like bands along $M$-$\Gamma$ path near the Fermi level. The band n = 5 possess a flat band just above Fermi level at $\Gamma$ point and an electron-like band at Fermi level along $M$-$\Gamma$ path. They contribute abundant electronic states to Fermi level. For these three predicted hydrides, their electronic density of states near the Fermi level is typically large and is dominated by the contribution from H atoms, which are very beneficial to the superconductivity.

To examine the chemical bonds of $Li_2NaH_{16}$, $Li_2NaH_{17}$ and $LiNa_3H_{23}$, we calculated electron localization function (ELF)[46], Bader charges[47] and crystal orbital Hamilton population (COHP)[48]. The ELF shows that there is enough charge localization between hydrogen atoms, indicating that H-H form covalent bonds (see Figure S5-7). The negative and positive COHP indicates bonding and antibonding, respectively. As shown in Figure S8-10, most H-H states below Fermi level also indicate H-H bonding interactions by COHP analysis. Bader charges analysis shows that electrons of Li and Na atoms transfer to hydrogen atoms. Moreover, Li and Na atoms transfer almost the same number of electrons about 0.78 $e\~$ per metal atom in $Li_2NaH_{16}$ and $Li_2NaH_{17}$. For $LiNa_3H_{23}$, Li and Na atoms transfer about 0.79 $e\~$ and 0.74 $e\~$ per atom, respectively. The transferred electrons occupy H-H antibonding orbitals which results the distance of H-H increases. It's noted that the shortest distance of H-H bond in these three compounds are above 1.00 Å, longer than $H_2$ gas molecule and a little longer than atomic metal hydrogen at 500 GPa[49]. Previous studies have shown that the existence of atomic hydrogen is beneficial to the increase of $T_c$.

Although $Li_2NaH_{17}$ has only one more H atom than $Li_2NaH_{16}$ in composition, its $T_c$ increases by nearly 120 K at 300 GPa. So, the introduced hydrogen atom in $Li_2NaH_{17}$ plays an important role in superconductivity. To gain a deep understanding of $T_c$ difference and elucidate the underlying mechanism of room-temperature superconductivity, we performed further analysis for electronic structures and Fermi surface nesting of $Li_2NaH_{16}$ and $Li_2NaH_{17}$. The projected H density of states at Fermi level increase from 0.4 States/eV/f.u. in $Li_2NaH_{16}$ to 0.63 States/eV/f.u. in $Li_2NaH_{17}$. There is also a change of van Hove singularity location. The introduction of an extra hydrogen atom can be seen as an electron doping in $Li_2NaH_{16}$, changing the electronic band structure of the system. It causes the Fermi level of $Li_2NaH_{17}$ shifts up compared to $Li_2NaH_{16}$ and bands above the Fermi level to shift downward. So, the number of bands crossing the Fermi level increases (see Figure 4). We could find the band (denoted n = 2) provides abundant H electronic states at Fermi level by analysis of FS sheets, resulting remarkable H density of electronic states and a van Hove singularity near the Fermi level. The appearance of van Hove singularity of light elements near the Fermi level contributes to the increase of $T_c$[50,51].

Interestingly, the band structure, DOS and FS sheets of thermodynamically stable $Li_2NaH_{17}$ are very similar with that of thermodynamically metastable $Li_2MgH_{16}$ at 300 GPa (see Figure S11). The $T_c$ of $Li_2MgH_{16}$ is 351 K at 300 GPa, and it's 340 K at 300 GPa for $Li_2NaH_{17}$. Their H-DOS contributes mainly to total DOS at the Fermi level. They have similar shapes of FS sheets and exist parallel Fermi surfaces in Brillouin zone, resulting strong and similar intensity of Fermi surface nesting (see Figure 3 and S20). Similar electronic structures and Fermi surface nesting result in similar $T_c$ at 300 GPa. Though $Li_2NaH_{16}$ also has a large H DOS at the Fermi level, it's intensity of Fermi surface nesting is much lower than $Li_2MgH_{16}$ and $Li_2NaH_{17}$. It results the $T_c$ of $Li_2NaH_{16}$ is much lower than $Li_2MgH_{16}$ and $Li_2NaH_{17}$. For the thermodynamically metastable $Rb_2MgH_{16}$[52] and thermodynamically stable $Li_2LaH_{17}$ and $Li_2YH_{17}$, their $T_c$s are much lower than room temperature. $Li_2YH_{17}$ and

$Rb_2MgH_{16}$ have small DOS at the Fermi level. Though $Li_2LaH_{17}$ possess strong Fermi surface nesting and van Hove singularity at the Fermi level, the large La-d DOS at the Fermi level suppresses its superconductivity.

We also analyzed the electronic structure of $LiNa_3H_{23}$, which is a thermodynamically stable type-I clathrate hydride with a room-temperature superconductivity. The projected DOS shows that H has a remarkable contribution near the Fermi level. Moreover, there are strong Fermi surface nesting near R point and along R-M path. Recent studies have also synthesized several type-I clathrate hydrides, such as $Ba_4H_{23}$[26], $La_4H_{23}$[27], $Lu_4H_{23}$[28] and $Eu_4H_{23}$[29]. Especially, $Ba_4H_{23}$ can be synthesized at 50 GPa and 1200 K, which was stable on decompression to 27 GPa. Compared to the type-I binary clathrate hydrides mentioned above, the $LiNa_3H_{23}$ has a better superconducting property, a room-temperature superconductor with a $T_c$ of 323 K at 320 GPa. It has a higher $T_c$ than all the reported type-I clathrate hydrides. It's possible to find high even room-temperature superconductors in type-I clathrate hydrides at moderate pressures.

**Conclusion**

We discovered several excellent room-temperature superconductors by carefully designing the calculation process to balance high temperature superconductivity and thermodynamic stability. Among them, $Li_2NaH_{17}$ and $LiNa_3H_{23}$ are predicted to be thermodynamically stable above 255 GPa and 335 GPa with $T_c$s above 300 K adopting typical type-I and type-II clathrate geometry, respectively. The high $T_c$ is attributed to the H-dominated DOS at the Fermi level and strong Fermi surface nesting. The recent synthesis of binary type-I hydrides $Ba_4H_{23}$, $La_4H_{23}$, $Lu_4H_{23}$ and $Eu_4H_{23}$ increase the confidence in synthesizing room-temperature superconductor $LiNa_3H_{23}$ at high pressure.

## Acknowledgements

This work was supported by the National Key Research and Development Program of China (No. 2022YFA1402304 and No. 2018YFA0305900), the National Natural Science Foundation of China (Grants No. 12122405, No. 52072188, No. 12274169, and No. 51632002), the Natural Sciences and Engineering Research Council of Canada (NSERC), program for Changjiang Scholars and Innovative Research Team in University (Grant No. IRT_15R23), and Jilin Provincial Science and Technology Development Project (20210509038RQ). Some of the calculations were performed at the High Performance Computing Center of Jilin University and using TianHe-1(A) at the National Supercomputer Center in Tianjin.

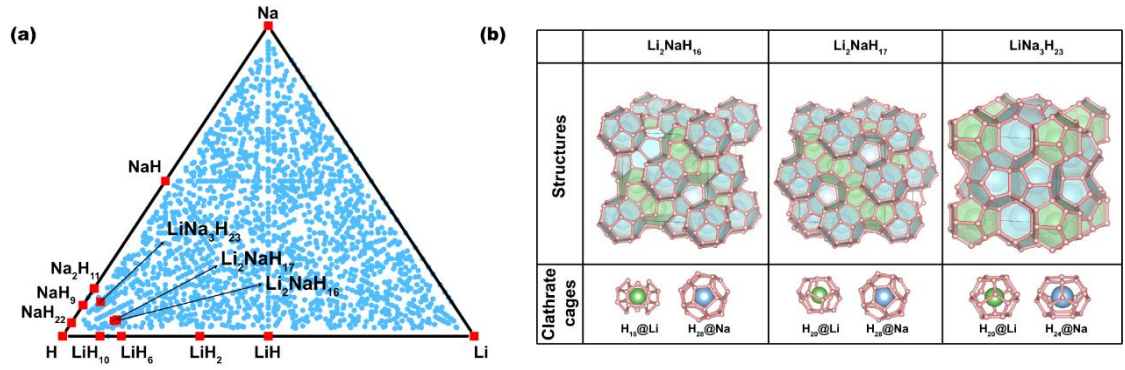

Figure 1 (a) The thermodynamically stable Li-Na-H ternaries relative to H[53], Li[54], Na[55], Li-H[56,57], Na-H[58,59] at 350 GPa. Red squares denote stable phases, blue dots denote non-stable phases. (b) Structures and clathrate cage structures of $Li_2NaH_{16}$, $Li_2NaH_{17}$ and $LiNa_3H_{23}$.

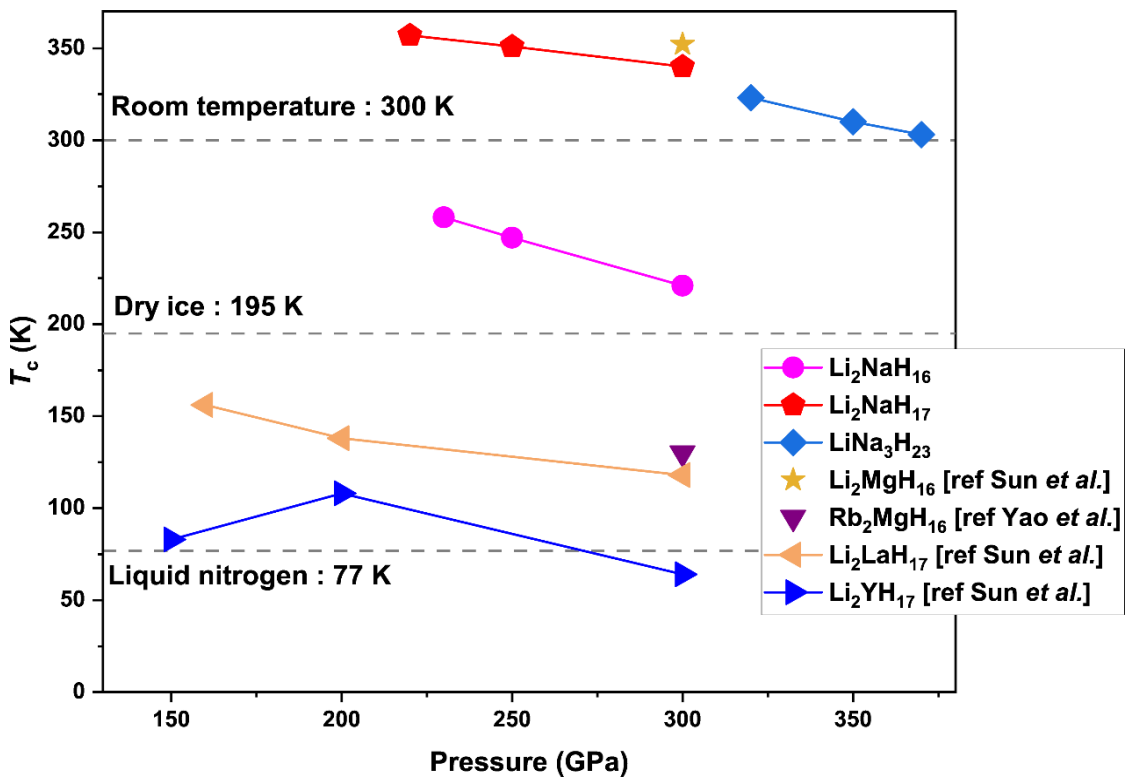

Figure 2 Superconducting critical temperatures with $\mu^* = 0.10$ as a function of pressure of $Li_2NaH_{16/17}$, $LiNa_3H_{23}$, $Li_2MgH_{16}$[20], $Rb_2MgH_{16}$[52] and $Li_2La/YH_{17}$[30].

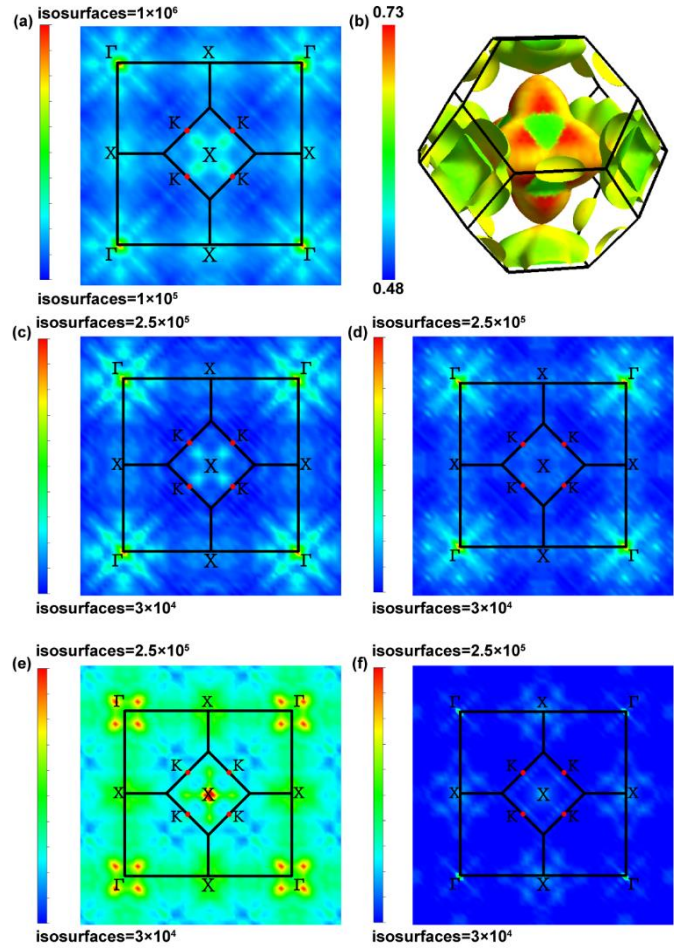

Figure 3 (a) 2D Fermi surface nesting function of Li$_2$NaH$_{17}$ in Brillouin zone. (b) 3D Fermi surface sheets of Li$_2$NaH$_{17}$ in Brillouin zone. (c) 2D Fermi surface nesting of Li$_2$NaH$_{17}$ from other bands to the band (n = 1) in Brillouin zone. (d) 2D Fermi surface nesting of Li$_2$NaH$_{17}$ from other bands to the band (n = 2) in Brillouin zone. (e) 2D Fermi surface nesting of Li$_2$NaH$_{17}$ from other bands to the band (n = 3) in Brillouin zone. (f) 2D Fermi surface nesting of Li$_2$NaH$_{17}$ from other bands to the band (n = 4) in Brillouin zone.

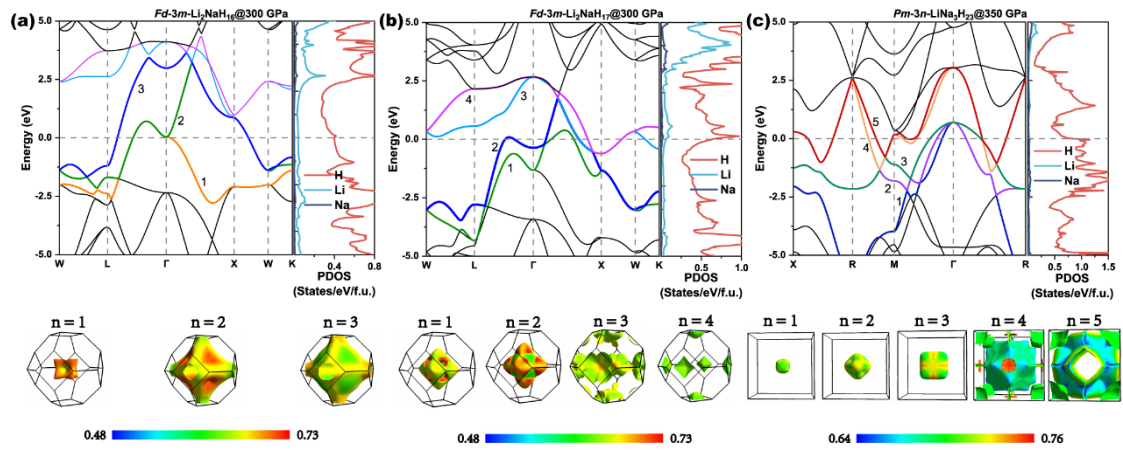

Figure 4 Calculated electronic band structures (left panel), projected DOS (middle panel) and FS sheets (right panel) of (a) $Fd$-$3m$-$Li_2NaH_{16}$ at 300 GPa, (b) $Fd$-$3m$-$Li_2NaH_{17}$ at 300 GPa, and (c) $Pm$-$3n$- $LiNa_3H_{23}$ at 350 GPa. The FS sheets were plotted using FermiSurfer software[60].

**References**

(1) Wigner, E.; Huntington, H. B. On the Possibility of a Metallic Modification of Hydrogen. *The Journal of Chemical Physics* **1935**, *3* (12), 764–770. https://doi.org/10.1063/1.1749590.

(2) McMahon, J. M.; Morales, M. A.; Pierleoni, C.; Ceperley, D. M. The Properties of Hydrogen and Helium under Extreme Conditions. *Rev. Mod. Phys.* **2012**, *84* (4), 1607–1653. https://doi.org/10.1103/RevModPhys.84.1607.

(3) Dias, R. P.; Silvera, I. F. Observation of the Wigner-Huntington Transition to Metallic Hydrogen. *Science* **2017**, *355* (6326), 715–718. https://doi.org/10.1126/science.aal1579.

(4) Eremets, M. I.; Drozdov, A. P.; Kong, P. P.; Wang, H. Semimetallic Molecular Hydrogen at Pressure above 350 GPa. *Nat. Phys.* **2019**, *15* (12), 1246–1249. https://doi.org/10.1038/s41567-019-0646-x.

(5) Loubeyre, P.; Occelli, F.; Dumas, P. Synchrotron Infrared Spectroscopic Evidence of the Probable Transition to Metal Hydrogen. *Nature* **2020**, *577* (7792), 631–635. https://doi.org/10.1038/s41586-019-1927-3.

(6) Ashcroft, N. W. Hydrogen Dominant Metallic Alloys: High Temperature Superconductors? *Phys. Rev. Lett.* **2004**, *92* (18), 187002. https://doi.org/10.1103/PhysRevLett.92.187002.

(7) Flores-Livas, J. A.; Boeri, L.; Sanna, A.; Profeta, G.; Arita, R.; Eremets, M. A Perspective on Conventional High-Temperature Superconductors at High Pressure: Methods and Materials. *Physics Reports* **2020**, *856*, 1–78. https://doi.org/10.1016/j.physrep.2020.02.003.

(8) Hilleke, K. P.; Zurek, E. Tuning Chemical Precompression: Theoretical Design and Crystal Chemistry of Novel Hydrides in the Quest for Warm and Light Superconductivity at Ambient Pressures. *Journal of Applied Physics* **2022**, *131* (7), 070901. https://doi.org/10.1063/5.0077748.

(9) Du, M.; Zhao, W.; Cui, T.; Duan, D. Compressed Superhydrides: The Road to Room Temperature Superconductivity. *J. Phys.: Condens. Matter* **2022**, *34* (17), 173001. https://doi.org/10.1088/1361-648X/ac4eaf.

(10) Duan, D.; Liu, Y.; Tian, F.; Li, D.; Huang, X.; Zhao, Z.; Yu, H.; Liu, B.; Tian, W.; Cui, T. Pressure-Induced Metallization of Dense $(H_2S)_2H_2$ with High-Tc Superconductivity. *Sci Rep* **2015**, *4* (1), 6968. https://doi.org/10.1038/srep06968.

(11) Drozdov, A. P.; Eremets, M. I.; Troyan, I. A.; Ksenofontov, V.; Shylin, S. I. Conventional Superconductivity at 203 Kelvin at High Pressures in the Sulfur Hydride System. *Nature* **2015**, *525* (7567), 73–76. https://doi.org/10.1038/nature14964.

(12) Li, Y.; Hao, J.; Liu, H.; Li, Y.; Ma, Y. The Metallization and Superconductivity of Dense Hydrogen Sulfide. *The Journal of Chemical Physics* **2014**, *140* (17), 174712. https://doi.org/10.1063/1.4874158.

(13) Peng, F.; Sun, Y.; Pickard, C. J.; Needs, R. J.; Wu, Q.; Ma, Y. Hydrogen Clathrate Structures in Rare Earth Hydrides at High Pressures: Possible Route to Room-Temperature Superconductivity. *Phys. Rev. Lett.* **2017**, *119* (10), 107001. https://doi.org/10.1103/PhysRevLett.119.107001.

(14) Liu, H.; Naumov, I. I.; Hoffmann, R.; Ashcroft, N. W.; Hemley, R. J. Potential High-$T_c$ Superconducting Lanthanum and Yttrium Hydrides at High Pressure. *Proc. Natl. Acad. Sci. U.S.A.* **2017**, *114* (27), 6990–6995. https://doi.org/10.1073/pnas.1704505114.

(15) Drozdov, A. P.; Kong, P. P.; Minkov, V. S.; Besedin, S. P.; Kuzovnikov, M. A.; Mozaffari, S.; Balicas, L.; Balakirev, F. F.; Graf, D. E.; Prakapenka, V. B.; Greenberg, E.; Knyazev, D. A.; Tkacz, M.; Eremets, M. I. Superconductivity at 250 K in Lanthanum Hydride under High Pressures. *Nature* **2019**, *569* (7757), 528–531. https://doi.org/10.1038/s41586-019-1201-8.

(16) Kong, P.; Minkov, V. S.; Kuzovnikov, M. A.; Drozdov, A. P.; Besedin, S. P.; Mozaffari, S.; Balicas, L.; Balakirev, F. F.; Prakapenka, V. B.; Chariton, S.; Knyazev, D. A.; Greenberg, E.; Eremets, M. I. Superconductivity up to 243 K in the Yttrium-Hydrogen System under High Pressure. *Nat Commun* **2021**, *12* (1), 5075. https://doi.org/10.1038/s41467-021-25372-2.

(17) Ma, L.; Wang, K.; Xie, Y.; Yang, X.; Wang, Y.; Zhou, M.; Liu, H.; Yu, X.; Zhao, Y.; Wang, H.; Liu, G.; Ma, Y. High-Temperature Superconducting Phase in Clathrate Calcium Hydride $CaH_6$ up to 215 K at a Pressure of 172 GPa. *Phys. Rev. Lett.* **2022**, *128* (16), 167001. https://doi.org/10.1103/PhysRevLett.128.167001.

(18) Wang, H.; Tse, J. S.; Tanaka, K.; Iitaka, T.; Ma, Y. Superconductive Sodalite-like Clathrate Calcium Hydride at High Pressures. *Proc. Natl. Acad. Sci. U.S.A.* **2012**, *109* (17), 6463–6466. https://doi.org/10.1073/pnas.1118168109.

(19) Li, Z.; He, X.; Zhang, C.; Wang, X.; Zhang, S.; Jia, Y.; Feng, S.; Lu, K.; Zhao, J.; Zhang, J.; Min, B.; Long, Y.; Yu, R.; Wang, L.; Ye, M.; Zhang, Z.; Prakapenka, V.; Chariton, S.; Ginsberg, P. A.; Bass, J.; Yuan, S.; Liu, H.; Jin, C. Superconductivity above 200 K Discovered in Superhydrides of Calcium. *Nat Commun* **2022**, *13* (1), 2863. https://doi.org/10.1038/s41467-022-30454-w.

(20) Sun, Y.; Lv, J.; Xie, Y.; Liu, H.; Ma, Y. Route to a Superconducting Phase above Room Temperature in Electron-Doped Hydride Compounds under High Pressure. *Phys. Rev. Lett.* **2019**, *123* (9), 097001. https://doi.org/10.1103/PhysRevLett.123.097001.

(21) Zhang, Z.; Cui, T.; Hutcheon, M. J.; Shipley, A. M.; Song, H.; Du, M.; Kresin, V. Z.; Duan, D.; Pickard, C. J.; Yao, Y. Design Principles for High-Temperature Superconductors with a Hydrogen-Based Alloy Backbone at Moderate Pressure. *Phys. Rev. Lett.* **2022**, *128* (4), 047001. https://doi.org/10.1103/PhysRevLett.128.047001.

(22) Zhong, X.; Sun, Y.; Iitaka, T.; Xu, M.; Liu, H.; Hemley, R. J.; Chen, C.; Ma, Y. Prediction of Above-Room-Temperature Superconductivity in Lanthanide/Actinide Extreme Superhydrides. *J. Am. Chem. Soc.* **2022**, *144* (29), 13394–13400. https://doi.org/10.1021/jacs.2c05834.

(23) Dolling, G.; Powell, B. M.; Sears, V. F. Neutron Diffraction Study of the Plastic Phases of Polycrystalline $SF_6$ and $CBr_4$. *Molecular Physics* **1979**, *37* (6), 1859–1883. https://doi.org/10.1080/00268977900101381.

(24) Wang 等。 - 2012 - Superconductive Sodalite-like Clathrate Calcium Hy.Pdf.

(25) Li, Y.; Hao, J.; Liu, H.; Tse, J. S.; Wang, Y.; Ma, Y. Pressure-Stabilized Superconductive Yttrium Hydrides. *Sci Rep* **2015**, *5* (1), 9948. https://doi.org/10.1038/srep09948.

(26) Peña-Alvarez, M.; Binns, J.; Martinez-Canales, M.; Monserrat, B.; Ackland, G. J.; Dalladay-Simpson, P.; Howie, R. T.; Pickard, C. J.; Gregoryanz, E. Synthesis of Weaire–Phelan Barium Polyhydride. *J. Phys. Chem. Lett.* **2021**, 7.

(27) Laniel, D.; Trybel, F.; Winkler, B.; Knoop, F.; Fedotenko, T.; Khandarkhaeva, S.; Aslandukova, A.; Meier, T.; Chariton, S.; Glazyrin, K.; Milman, V.; Prakapenka, V.; Abrikosov, I. A.; Dubrovinsky, L.; Dubrovinskaia, N. High-Pressure Synthesis of Seven Lanthanum Hydrides with a Significant Variability of Hydrogen Content. *Nat Commun* **2022**, *13* (1), 6987. https://doi.org/10.1038/s41467-022-34755-y.

(28) Li, Z.; He, X.; Zhang, C.; Lu, K.; Bin, B.; Zhang, J.; Zhang, S.; Zhao, J.; Shi, L.; Feng, S.; Wang, X.; Peng, Y.; Yu, R.; Wang, L.; Li, Y.; Bass, J.; Prakapenka, V.; Chariton, S.; Liu, H.; Jin, C. Superconductivity above 70 K Experimentally Discovered in Lutetium Polyhydride. arXiv March 9, 2023. http://arxiv.org/abs/2303.05117 (accessed 2023-03-16).

(29) Semenok, D. V.; Zhou, D.; Kvashnin, A. G.; Huang, X.; Galasso, M.; Kruglov, I. A.; Ivanova, A. G.; Gavriliuk, A. G.; Chen, W.; Tkachenko, N. V.; Boldyrev, A. I.; Troyan, I.; Oganov, A. R.; Cui, T. Novel Strongly Correlated Europium Superhydrides. *J. Phys. Chem. Lett.* **2021**, *12* (1), 32–40. https://doi.org/10.1021/acs.jpclett.0c03331.

(30) Sun, Y.; Wang, Y.; Zhong, X.; Xie, Y.; Liu, H. High-Temperature Superconducting Ternary Li − R − H Superhydrides at High Pressures ( R = Sc , Y , La ). *Phys. Rev. B* **2022**, *106* (2), 024519. https://doi.org/10.1103/PhysRevB.106.024519.

(31) Kasper, J. S.; Hagenmuller, P.; Pouchard, M.; Cros, C. Clathrate Structure of Silicon Na8Si46 and NaxSi136 (x < 11). *Science* **1965**, *150* (3704), 1713–1714. https://doi.org/10.1126/science.150.3704.1713.

(32) Frank, F. C.; Kasper, J. S. Complex Alloy Structures Regarded as Sphere Packings. II. Analysis and Classification of Representative Structures. *Acta Cryst* **1959**, *12* (7), 483–499. https://doi.org/10.1107/S0365110X59001499.

(33) San-Miguel, A.; Mélinon, P.; Connétable, D.; Blase, X.; Tournus, F.; Reny, E.; Yamanaka, S.; Itié, J. P. Pressure Stability and Low Compressibility of Intercalated Cagelike Materials: The Case of Silicon Clathrates. *Phys. Rev. B* **2002**, *65* (5), 054109. https://doi.org/10.1103/PhysRevB.65.054109.

(34) Zhu, L.; Borstad, G. M.; Liu, H.; Guńka, P. A.; Guerette, M.; Dolyniuk, J.-A.; Meng, Y.; Greenberg, E.; Prakapenka, V. B.; Chaloux, B. L.; Epshteyn, A.; Cohen, R. E.; Strobel, T. A. Carbon-Boron Clathrates as a New Class of $Sp^3$-Bonded Framework Materials. *Sci. Adv.* **2020**, *6* (2), eaay8361. https://doi.org/10.1126/sciadv.aay8361.

(35) Ding, H.-B.; Feng, Y.-J.; Jiang, M.-J.; Tian, H.-L.; Zhong, G.-H.; Yang, C.-L.; Chen, X.-J.; Lin, H.-Q. Ambient-Pressure High- T c Superconductivity in Doped Boron-Nitrogen Clathrates La ( BN ) 5 and Y ( BN ) 5. *Phys. Rev. B* **2022**, *106* (10), 104508. https://doi.org/10.1103/PhysRevB.106.104508.

(36) Heyrovska, R. Atomic, Ionic and Bohr Radii Linked via the Golden Ratio for Elements of Groups 1 - 8 Including Lanthanides and Actinides.

(37) Semenok, D. V.; Troyan, I. A.; Ivanova, A. G.; Kvashnin, A. G.; Kruglov, I. A.; Hanfland, M.; Sadakov, A. V.; Sobolevskiy, O. A.; Pervakov, K. S.; Lyubutin, I. S.; Glazyrin, K. V.; Giordano, N.; Karimov, D. N.; Vasiliev, A. L.; Akashi, R.; Pudalov, V. M.; Oganov, A. R.

Superconductivity at 253 K in Lanthanum–Yttrium Ternary Hydrides. *Materials Today* **2021**, *48*, 18–28. https://doi.org/10.1016/j.mattod.2021.03.025.

(38) Bi, J.; Nakamoto, Y.; Zhang, P.; Shimizu, K.; Zou, B.; Liu, H.; Zhou, M.; Liu, G.; Wang, H.; Ma, Y. Giant Enhancement of Superconducting Critical Temperature in Substitutional Alloy (La,Ce)H9. *Nat Commun* **2022**, *13* (1), 5952. https://doi.org/10.1038/s41467-022-33743-6.

(39) Chen, W.; Huang, X.; Semenok, D. V.; Chen, S.; Zhang, K.; Oganov, A. R.; Cui, T. Enhancement of the Superconducting Critical Temperature Realized in the La-Ce-H System at Moderate Pressures. **2022**. https://doi.org/10.48550/ARXIV.2203.14353.

(40) Huang, G.; Luo, T.; Dalladay-Simpson, P.; Chen, L.-C.; Cao, Z.-Y.; Peng, D.; Gorelli, F. A.; Zhong, G.-H.; Lin, H.-Q.; Chen, X.-J. Synthesis of Superconducting Phase of $La_{0.5}Ce_{0.5}H_{10}$ at High Pressures. **2022**. https://doi.org/10.48550/ARXIV.2208.05199.

(41) Kawaji, H.; Horie, H.; Yamanaka, S.; Ishikawa, M. Superconductivity in the Silicon Clathrate Compound $(Na,Ba)_xSi_{46}$. *Phys. Rev. Lett.* **1995**, *74* (8), 1427–1429. https://doi.org/10.1103/PhysRevLett.74.1427.

(42) Yamanaka, S.; Enishi, E.; Fukuoka, H.; Yasukawa, M. High-Pressure Synthesis of a New Silicon Clathrate Superconductor, $Ba_8Si_{46}$. *Inorg. Chem.* **2000**, *39* (1), 56–58. https://doi.org/10.1021/ic990778p.

(43) Togo, A.; Tanaka, I. First Principles Phonon Calculations in Materials Science. *Scripta Materialia* **2015**, *108*, 1–5. https://doi.org/10.1016/j.scriptamat.2015.07.021.

(44) Sanna, A.; Flores-Livas, J. A.; Davydov, A.; Profeta, G.; Dewhurst, K.; Sharma, S.; Gross, E. K. U. Ab Initio Eliashberg Theory: Making Genuine Predictions of Superconducting Features. *J. Phys. Soc. Jpn.* **2018**, *87* (4), 041012. https://doi.org/10.7566/JPSJ.87.041012.

(45) Tse, J. S.; Yao, Y.; Tanaka, K. Novel Superconductivity in Metallic $SnH_4$ under High Pressure. *Phys. Rev. Lett.* **2007**, *98* (11), 117004. https://doi.org/10.1103/PhysRevLett.98.117004.

(46) Becke, A. D.; Edgecombe, K. E. A Simple Measure of Electron Localization in Atomic and Molecular Systems. *The Journal of Chemical Physics* **1990**, *92* (9), 5397–5403. https://doi.org/10.1063/1.458517.

(47) Tang, W.; Sanville, E.; Henkelman, G. A Grid-Based Bader Analysis Algorithm without Lattice Bias. *J. Phys.: Condens. Matter* **2009**, *21* (8), 084204. https://doi.org/10.1088/0953-8984/21/8/084204.

(48) Deringer, V. L.; Tchougréeff, A. L.; Dronskowski, R. Crystal Orbital Hamilton Population (COHP) Analysis As Projected from Plane-Wave Basis Sets. *J. Phys. Chem. A* **2011**, *115* (21), 5461–5466. https://doi.org/10.1021/jp202489s.

(49) Labet, V.; Gonzalez-Morelos, P.; Hoffmann, R.; Ashcroft, N. W. A Fresh Look at Dense Hydrogen under Pressure. I. An Introduction to the Problem, and an Index Probing Equalization of H–H Distances. *The Journal of Chemical Physics* **2012**, *136* (7), 074501. https://doi.org/10.1063/1.3679662.

(50) Liu, X.; Huang, X.; Song, P.; Wang, C.; Zhang, L.; Lv, P.; Liu, L.; Zhang, W.; Cho, J.-H.; Jia, Y. Strong Electron-Phonon Coupling Superconductivity in Compressed $\alpha-MoB_2$

Induced by Double Van Hove Singularities. *Phys. Rev. B* **2022**, *106* (6), 064507. https://doi.org/10.1103/PhysRevB.106.064507.

(51) Gai, T.-T.; Guo, P.-J.; Yang, H.-C.; Gao, Y.; Gao, M.; Lu, Z.-Y. Van Hove Singularity Induced Phonon-Mediated Superconductivity above 77 K in Hole-Doped SrB$_3$C$_3$. *Phys. Rev. B* **2022**, *105* (22), 224514. https://doi.org/10.1103/PhysRevB.105.224514.

(52) Yao, S.; Song, Q.; Hu, W.; Wang, D.; Peng, L.; Shi, T.; Chen, J.; Liu, X.; Lin, J.; Chen, X. Fermi Surface Topology and Anisotropic Superconducting Gap in Electron-Doped Hydride Compounds at High Pressure. *Phys. Rev. Materials* **2022**, *6* (3), 034801. https://doi.org/10.1103/PhysRevMaterials.6.034801.

(53) Pickard, C. J.; Needs, R. J. Structure of Phase III of Solid Hydrogen. *Nature Phys* **2007**, *3* (7), 473–476. https://doi.org/10.1038/nphys625.

(54) Lv, J.; Wang, Y.; Zhu, L.; Ma, Y. Predicted Novel High-Pressure Phases of Lithium. *Phys. Rev. Lett.* **2011**, *106* (1), 015503. https://doi.org/10.1103/PhysRevLett.106.015503.

(55) Ma, Y.; Eremets, M.; Oganov, A. R.; Xie, Y.; Trojan, I.; Medvedev, S.; Lyakhov, A. O.; Valle, M.; Prakapenka, V. Transparent Dense Sodium. *Nature* **2009**, *458* (7235), 182–185. https://doi.org/10.1038/nature07786.

(56) Zurek, E.; Hoffmann, R.; Ashcroft, N. W.; Oganov, A. R.; Lyakhov, A. O. A Little Bit of Lithium Does a Lot for Hydrogen. *Proc. Natl. Acad. Sci. U.S.A.* **2009**, *106* (42), 17640–17643. https://doi.org/10.1073/pnas.0908262106.

(57) Chen, Y.; Geng, H. Y.; Yan, X.; Sun, Y.; Wu, Q.; Chen, X. Prediction of Stable Ground-State Lithium Polyhydrides under High Pressures. *Inorg. Chem.* **2017**, *56* (7), 3867–3874. https://doi.org/10.1021/acs.inorgchem.6b02709.

(58) Baettig, P.; Zurek, E. Pressure-Stabilized Sodium Polyhydrides: NaH$_n$ ( n > 1 ). *Phys. Rev. Lett.* **2011**, *106* (23), 237002. https://doi.org/10.1103/PhysRevLett.106.237002.

(59) Shipley, A. M.; Hutcheon, M. J.; Needs, R. J.; Pickard, C. J. High-Throughput Discovery of High-Temperature Conventional Superconductors. *Phys. Rev. B* **2021**, *104* (5), 054501. https://doi.org/10.1103/PhysRevB.104.054501.

(60) Kawamura, M. FermiSurfer: Fermi-Surface Viewer Providing Multiple Representation Schemes. *Computer Physics Communications* **2019**, *239*, 197–203. https://doi.org/10.1016/j.cpc.2019.01.017.